\documentclass[preprintnumbers,showpacs,amsmath,amssymb,nofootinbib,twocolumn]{revtex4-1}
\usepackage[T1]{fontenc}
\usepackage{graphicx}
\usepackage{epsfig}
\usepackage{dcolumn}
\usepackage{multirow}
\usepackage{rotating}
\usepackage{verbatim}
\usepackage{bm}
\usepackage{color}
\usepackage{soul}
\usepackage{float}
\usepackage{hyperref}
\usepackage[utf8]{inputenc}

\def\lsim{\raise0.3ex\hbox{$<$\kern-0.75em\raise-1.1ex\hbox{$\sim$}}}
\def\gsim{\raise0.3ex\hbox{$>$\kern-0.75em\raise-1.1ex\hbox{$\sim$}}}

\def\pom{{I\!\!P}}

\newcommand{\rr}{\mbox{\boldmath $r$}}

\def\beqa{\begin{eqnarray}}
\def\eeqa{\end{eqnarray}}
\newcommand\kzero{{\rm K}_0}
\newcommand\kone{{\rm K}_1}

\begin{document}

\title{Soft diffraction within the QCD color dipole picture}
\author{G.M. Peccini} 
\email{guilherme.peccini@ufrgs.br}
\author{F. Kopp}
\email{fabio.kopp@ufrgs.br}
\author{ M.V.T. Machado}
\email{magnus@if.ufrgs.br}
\affiliation{High Energy Physics Phenomenology Group, GFPAE. Institute of Physics, Federal University of Rio Grande do Sul (UFRGS)\\
Caixa Postal 15051, CEP 91501-970, Porto Alegre, RS, Brazil} 
\author{D.A. Fagundes}
\email{daniel.fagundes@ufsc.br}
\affiliation{Department of Exact Sciences and Education, CEE. Federal University
of Santa Catarina (UFSC) - Blumenau Campus, 89065-300, Blumenau, SC, Brazil}

\begin{abstract}

In this work we consider the QCD parton saturation models to describe soft interactions at the high energy limit.  The total and elastic cross sections as well as the elastic slope parameter are obtained for proton-proton and pion-proton collisions and compared to recent 
experimental results. The analyses are done within the color dipole formalism taking into account saturation models which have been tested against DIS data. The main point is that the match between soft and hard interaction occurs in the saturation region which can be described by high density QCD approaches. Discussion is performed on the main theoretical uncertainties associated with calculations. 

\end{abstract}

\maketitle
\tableofcontents

\section{Introduction} 

Describing soft processes using the QCD degrees of freedom is a quite difficult task as they are dominated by long distance (non-perturbative) physics. It has been shown that soft observables as the total and elastic cross sections depend on the transition region between the high parton density system (saturation domain) and perturbative QCD region \cite{Bartels:2002uf,Carvalho:2007cf,Arguelles:2015wba}. The parton saturation phenomenon \cite{Gelis:2010nm,Weigert:2005us,JalilianMarian:2005jf} is a well established property of high energy systems and gives a high quality description of inclusive and exclusive Deep Inelastic Scattering (DIS) data. As evidences for the successfulness of such approach we quote the description of light meson photoproduction cross section \cite{Caldwell:2001ky,Kowalski:2003hm,Forshaw:2003ki,Marquet:2007qa,Kowalski:2006hc,Armesto:2014sma} and diffractive DIS (DDIS) \cite{Forshaw:2006np,GolecBiernat:1999qd}. Both are hard processes in which an important contribution to cross section comes from kinematic region in the vicinity of the saturation momentum, $Q_s$. This dimensional scale increases in the high energy region. A well known formalism that is intuitive and where saturation physics can be easily implemented is the QCD color dipole picture. A seminal work on this approach is Ref. \cite{Nikolaev:1993th} where the unitarity corrections to the proton structure function at small-$x$ were derived and predictions are done for DDIS and nuclear shadowing. There it was demonstrated that the factorization of the photon-induced cross sections between the Fock states wave functions of photon and  multiparticle dipole cross section provides clear identification of the partial waves of the dipole cross section as an object of the s-channel unitarization. Moreover, applications to the soft hadronic scattering within the same formalism has been done in \cite{Nikolaev:1993ke}. It is expected \cite{Bartels:2002uf} that soft processes measured for instance at the Large Hadron Collider (LHC) in hadron-hadron collisions probe distances about $r\sim 1/Q_s\ll R_h$, with $R_h$ being the hadron radius. In this context, hadron scattering at the LHC could be described by color dipoles as the correct degrees of freedom even at large transverse distances. Moreover, it has been shown that the cross sections for soft hadron-hadron collisions within saturation approaches satisfy the Froissart-Martin bound \cite{Carvalho:2007cf,Arguelles:2015wba}. In addition, also the role played by the unitarized hard Pomeron contribution to the soft observables has been carefully discussed in Refs. \cite{Cudell:2008zz,Cudell:2008yb}. The relationship and equivalence between the BFKL and dipole equation kernels are investigated by means of explicit calculations in light-cone perturbation theory. A dipole equation, equivalent to the usual equation for interactions between four reggeized gluons, is given in the large $N_c$ limit. The leading trajectory of the four-gluon system is bounded by $2 \alpha _{I\!P}$ $-$ $ 1$ with $\alpha _{I\!P}$ being the BFKL pomeron intercept. \cite{Chen:1995pa} 

An important property of the saturation formalism is the geometric scaling phenomenon \cite{Stasto:2000er}, which means that the scattering amplitude and corresponding cross sections can scale on the dimensionless scale $\tau=\mu^ 2/Q_s^2$, where $\mu^2$ is the typical hard scale in the scattering process. For instance, $\mu^2=Q^2$ is the photon virtuality in DIS (i.e., the nucleon structure functions $F_2,\,F_L$) and Deeply Virtual Compton Scattering (DVCS) processes or $\mu^2=(Q^2+M_V^2)$ in case of exclusive electroproduction ($Q^2\neq 0$) and photoproduction ($Q^2= 0$) of vector mesons of masses $M_V$. The treatment of vector meson production (including production of their excited states) within the color dipole picture in terms of the scanning radius, $r_S$, was first addressed in Ref. \cite{Kopeliovich:1993gk} with the identification of the relevant hard scale $(Q^2+M_V^2)$ (for a comprehensive and pedagogical review about vector mesons we quote Ref. \cite{Ivanov:2004ax}, where experimental results summarized and theoretical formalisms are compared with emphasis on the BFKL color dipole and kt-factorization approaches).   Deviations from geometric scaling are also known when the system is far from the saturation domain. Geometric scaling is an intrinsic property of non-linear QCD evolution equations \cite{Gelis:2010nm,Weigert:2005us,JalilianMarian:2005jf} in the asymptotic energy regime, $\sqrt{s}\rightarrow \infty$. This scaling property has been used in recent years to construct phenomenological models for the QCD dynamics at high energies. A very intuitive picture of inclusive or exclusive DIS process is the color dipole picture \cite{Nikolaev:1990ja,Nikolaev:1991et,Mueller:1993rr,Mueller:1994jq}.  In that picture the Deep Inelastic Scattering process can be seen as a succession in time of three factorizable subprocesses: i) 
the photon fluctuates in a quark-anti-quark pair with transverse separation $r\sim 1/Q$ long after the interaction, ii) this 
color dipole interacts with the nucleon target, iii) the quark pair is projected into the considered final state. The nucleon structure function is related to the $\gamma^* p $ cross section as $F_{2}(x,Q^{2})= \frac{Q^2}{4\pi^2 \alpha_{em}} \sigma_{tot}^{\gamma^* p}$. The latter is the overlap of the dipole cross section on the transverse and longitudinal  photon wave-functions. The interaction is then factorized in the simple formulation \cite{Nikolaev:1990ja,Nikolaev:1991et,Mueller:1993rr,Mueller:1994jq},
\begin{eqnarray}
\sigma_{tot}^{\gamma^*p}(W_{\gamma p},Q^2)  & = &  \int dz \,d^2\rr\left(|\Psi_{T}\,(z,\,\rr)|^2+ |\Psi_{L}\,(z,\,\rr)|^2\right) \nonumber \\
&\times & \sigma_{dip}\,(\tilde{x},\,\rr),
\label{F2dip}
\end{eqnarray}
where $z$ is the longitudinal momentum fraction of the quark in the color dipole, $\tilde{x} = \frac{ Q^2+m_q^2}{W_{\gamma^{*} p}^2+Q^2}$ is equivalent to the Bjorken variable and provides an interpolation for the $Q^2\rightarrow 0$ limit. The mass of the quark of flavor $f$ is labeled as $m_f$. The photon wave-functions $\Psi_{T,L}(\rr,z;Q^2)$ are determined from light cone perturbation theory and the dipole hadron cross section $\sigma_{dip}(x,\rr)=2\int d^2 b\,N_{dip}(x,\rr,b)$  contains all the information about the target and the strong interaction physics (including the impact parameter, $b$, dependence). As an example, the celebrated GBW parameterization \cite{GolecBiernat:1998js,GolecBiernat:1999qd} takes the eikonal-like form,
\begin{eqnarray}
\sigma_{dip} \,({x}, \,\rr^2)  & = &  \sigma_0 \, \left[\, 1- \exp
\left(-\frac{\rr^2Q_{s}^2}{4} \right)^{\gamma_s} \right],\\
 Q_{s}^2\,(x)  & = &  \left( \frac{x_0}{\tilde{x}}
\right)^{\lambda} \,\,\mathrm{GeV}^2\,. \label{gbwdip}
\end{eqnarray}
where $Q_s$ is the saturation scale. The parameters are obtained from  fit to the HERA data producing $\sigma_0=27.43$ mb, $\lambda=0.248$ and $x_0=0.40\cdot 10^{-4}$ for a 5-flavor analysis (See Ref. ~\cite{Golec-Biernat:2017lfv} for an updated fitting procedure). Here, additional parameters are the effective light quark mass, $m_f=0.14$
GeV, which plays the role of a regulator at the photoproduction limit. The charm (bottom) mass is set to be $m_c=1.4\,(4.6)$ GeV. The GBW parameterization presents a geometric scaling form, $\sigma_{dip}\propto f(\rr^2Q_{\mathrm{sat}}^2)$. For small dipoles, $\rr^2\leq 1/Q_{\mathrm{sat}}^2$, it can be approximated by $\sigma_{dip} \simeq \sigma_0 (\rr^2Q_{\mathrm{sat}}^2/4)$, where the effective anomalous dimension is equal to one, $\gamma_s =1$. 

The advent of the LHC opened a new window for the studies on diffraction, elastic and inelastic scattering as they are not strongly contaminated by non-diffractive events. This is translated in the Regge theory language saying that the  scattering amplitude is completely determined by a Pomeron exchange. The current measurements on these soft observables at the LHC in proton-proton collisions span a wide range energies from 100 GeV including the very recent LHC data at 2.76, 7.0, 8.0  and 13 TeV \cite{Tanabashi:2018oca,Nemes:2018tk,Antchev:2017dia,Antchev:2017yns,Antchev:2016vpy,Aaboud:2016ijx,Aad:2014dca,Antchev:2015zza,Antchev:2013iaa}. In the context of saturation physics the soft Pomeron may be understood as a unitarized perturbation Pomeron \cite{Motyka:2003bn}. It can be shown that the trajectory of the soft Pomeron could emerge as a result of the  interplay between perturbative physics of the hard Pomeron and the confining properties of the QCD vacuum. Specifically, local unitarization in the impact parameter plane can lead to a reasonable description of intercept and the slope of soft Pomeron \cite{Motyka:2003bn}. In the present work, we investigate the soft observable in the small-$t$ regime within the color  dipole picture and parton saturation approaches.  Of course, some words of reservation are needed here. We are aware that saturation scale in general  is relatively small and its role in perturbative QCD (pQCD) is highly debatable. The issue of extension of the color dipole language from the hard BFKL  pQCD region of small dipoles to the soft pomeron at hadronic scales  remains open(we quote Ref.  \cite{{Fiore:2012yi}} and references therein for a careful discussion about this problem). 

The paper is organized as follows. In the next section we summarize the theoretical information to compute the cross section for hadron-hadron collisions in two color dipole approaches. First, we consider the asymptotic cross section following Ref. \cite{Arguelles:2015wba}, where the $pp$ cross section is assumed to be dominated by two-gluon production in the final state, $pp\rightarrow gg+X$. There, the main ingredients are the gluon distribution of the projectile and the partonic cross section associated to the interaction $gN\rightarrow gg+X$. We also consider the model presented in Ref. \cite{Bartels:2002uf}, where the cross section for hadron-proton collision is viewed in a similar way as Eq. (\ref{F2dip}), where the virtual photon wave-function is replaced by the corresponding wave function for the hadron projectile. The hadron-proton interaction is computed using the dipole-proton amplitude constrained by DIS data. The numerical results from both models are compared to experimental measurements focusing in the LHC kinematic regime.  In the last section we discuss the main theoretical uncertainties and present the main conclusions.

\section{Color Dipole Models}

\subsection{Asymptotic model}

Our first investigation will consider the color dipole approach applied to hadron-hadron collisions proposed in Ref. \cite{Arguelles:2015wba}. For simplicity, we address initially the case for proton-proton collisions in colliders. The formalism is able to provide us the production cross section of (heavy or light) quark pairs or gluons at the final state. Namely, similarly to
photon-hadron interactions, the total quark production cross
section is given by \cite{Nikolaev:1995ty,Kopeliovich:2002yv},
\begin{eqnarray}
\sigma(pp\to q\bar q X) & = &
2\int_0^{-\ln(\frac{2m_q}{\sqrt{s}})}dy\,
x_1G\left(x_1,\mu_F^2\right) \nonumber \\
& \times &    \sigma(GN\to q\bar q X)\,\,,
\label{ccppdip}
\end{eqnarray}
where  $y=\frac{1}{2}\ln(x_1/x_2)$ is the rapidity of the pair,
$\mu_F\sim m_Q$ is the factorization scale. The quantity $x_1G(x_1,\mu_F^2)$ is
the projectile gluon density at scale $\mu_F$ and the partonic cross
section $\sigma(GN\to q\bar q X)$ is given by \cite{Nikolaev:1995ty}, 
\begin{eqnarray}
\label{eq:all}
\sigma(GN\to q\bar q X) &=&
\int dz \, d^2 \rr \left|\Psi_{G\to q\bar
q}(z,\rr)\right|^2 \nonumber \\
& \times & \sigma_{q\bar q G}(z,\rr)\,\,, 
\end{eqnarray}
 with
$\Psi_{G\to q\bar q}$ being the pQCD calculated distribution
amplitude , which describes the dependence of the $|q \bar q
\rangle$ Fock component on transverse separation and fractional
momentum. It is given by,
\begin{eqnarray}
\label{eq:lcwf}
\left|\Psi_{G\to q\bar q}(z,\rr)\right|^2 & = & 
\frac{\alpha_s(\mu_R)}{(2\pi)^2}\left\{m_q^2\kzero^2(m_q r) \right. \nonumber \\
&+ & \left. \left[z^2+{(1-z)}^2\right]m_q^2\kone^2(m_q r)\right\},
\end{eqnarray}
where $\alpha_s(\mu_R)$ is the strong
coupling constant, which is probed at a renormalization scale
$\mu_R\sim m_Q$. We notice that the wavefunction will lead to a dominance of dipole sizes around $r\sim 1/m_q$ in the corresponding $r$-integration. Therefore, for heavy quark production the color transparency behavior from dipole cross section, $\sigma_{dip}(r)\propto r^ 2$, will be the main contribution (pQCD). In the charm case, an important contribution should come from saturation region as the typical dipole size, $r\simeq 1$ GeV$^{-1}$, can reach order of magnitude similar to the  saturation radius, $R_s(x)=1/Q_s(x)\propto (\sqrt{s})^{-\lambda/2}$ (with $\lambda\simeq 0.3$). On the other hand, for light quarks, $m_q\simeq 0.14$ GeV, we are deep in the parton saturation (very low-$x_2$ and small scale of probe) and non-perturbative regions. This will be the case in the following calculation.

In the partonic cross section, $\sigma_{q\bar qG}$ is the cross section
for scattering a color neutral quark-antiquark-gluon system on the
target and is directly related with the dipole cross section as
follows
\begin{equation}
\label{eq:qqG}
\sigma_{q\bar qG}=\frac{9}{8}\left[\sigma_{dip}(x_2,z \rr)
+\sigma_{dip}(x_2,\bar{z}\rr)\right]
-\frac{1}{8}\sigma_{dip}(x_2,\rr) . 
\end{equation}
The equation above was first derived in Ref. \cite{Nikolaev:1993th} and  the main idea is that at high energies a gluon $G$ from the hadron projectile  can develop a fluctuation which contains a $Q\bar Q$ pair. Interaction with the color field of the
target then may release these heavy quarks. Such an approach is valid
for high energies, where the coherence length $l_c \approx 1/x_2$ is
larger than the target radius. Hence, it is a natural framework to include the parton saturation effects and to make use of the fact that the dipole
cross section is universal, i.e., it is processed independently. For sake of completeness, the parton momentum fractions are written in terms of quark pair rapidity and masses, $x_{1,2} = \frac{2m_Q}{\sqrt{s}}\exp(\pm y)$.

Following Ref. \cite{Arguelles:2015wba}, we obtain the asymptotic hadron-hadron total cross section within the color dipole approach considering the dominant process, $pp\rightarrow GGX$, at high energies. Now, the gluon $G$ from the projectile hadron develops a fluctuation which contains a two-gluon ($GG$) pair which further interacts with the target's color field.  Accordingly, the expression for the total cross section for gluon production at final state is given by \cite{Nikolaev:2005zj},
\begin{equation}
\sigma_{pp\to GG X}  = 
2\int_0^{\tilde{y}}dy\,
x_1G\left(x_1,\mu_F^2\right)\sigma(GN\to GGX), 
\label{ccppdip}
\end{equation}
where $\tilde{y}=-\ln\left(\frac{2m_G}{\sqrt{s}}\right)$ and the effective gluon mass, $m_G$, has been introduced in order to regularize the calculation. Thus, in this case one has $x_{1,2} = \frac{2m_G}{\sqrt{s}}\exp(\pm y)$.

The new partonic cross section $\sigma_{GN\to GG X}$ is given by, 
\begin{equation}
\label{eq:partgg}
\sigma_{GN\to GG X}=\int dz \, d^2 \rr \left|\Psi_{G\to GG}(z,\rr)\right|^2 \sigma_{GGG}(z,\rr), 
\end{equation}
with
$\Psi_{G\to GG}$ being the corresponding distribution
amplitude associated to the $|GG\rangle$ Fock state. It is obtained from Eq. (\ref{eq:lcwf}) in the following way, $|\Psi_{G\to GG}|^ 2=2(N_c-1)|\Psi_{G\to q \bar q}|^2$ .  The partonic cross section, $\sigma_{GGG}$, is the cross section
for scattering a color neutral three gluon system on the
target and is directly related with the dipole cross section as follows \cite{Nikolaev:2005zj},
 \begin{equation}
\label{eq:GGG}
 \sigma_{GGG}  = \frac{1}{2}\left[\sigma_{dip}(x_2,z \rr)  
+ \sigma_{dip}(x_2,\bar{z}\rr)+\sigma_{dip}(x_2,\rr)\right ]. 
 \end{equation}
The approach described above is derived from the nonlinear $k_{\perp}$-factorization approach for the production of hard gluon-gluon dijets in gluon-hadron collisions when the coherence condition holds. This gluon-gluon dijet cross section can be investigated in different color representations and their classification in universality classes can be defined. 

Now, we will present the corresponding phenomenology using Eq. (\ref{ccppdip}). From Ref. \cite{Arguelles:2015wba}, basically we identify two main shortcomings: the very low value for the effective gluon mass, $m_G = 154\,\mathrm{MeV}\, <\Lambda_{QCD}$, and the identification of the scale $\mu$ with the starting evolution scale in the gluon PDFs considered, $\mu^ 2=Q_0^2$. Here, we will use the value $m_G=400$ MeV that is consistent with the usual values in Refs. \cite{Fagundes:2011zx,Bahia:2015hha,Broilo:2019yuo}. Moreover, the gluon PDF probed in the low scale $\mu^2=m_G^2=0.16$ GeV$^2$ will be given by a prediction from the parton saturation physics,
\begin{eqnarray}
\label{gluon1}
x\, G(x,Q^2) = \frac{3\,\sigma_0Q_{s}^2}{4 \pi^2 \alpha_s}\;
\left[1-\left(1+\frac{Q^2}{Q_{s}^2}\right)\;e^{-\frac{Q^2}{Q_{s}^2}}\right],
\end{eqnarray}
where updated values for the GBW model parameters have been used \cite{Golec-Biernat:2017lfv}. Consistently, for the dipole cross section we have used the GBW parametrization. It should be stressed that the result is parameter free and corresponds to the soft Pomeron contribution to the cross section.

Let us discuss quantitatively the main ingredients in the asymptotic model. For example, take the LHC energy of $\sqrt{s}=13$ TeV ($\tilde{y}=-\ln(2m_G/\sqrt{s})=9.7$ and $\Delta y=2\tilde{y}\simeq 19$). At central gluon rapidity, $y=0$, the longitudinal momentum fractions will be $x_1=x_2=2m_G/\sqrt{s}\approx 6\times 10^{-5}$ whereas at very forward rapidity $x_2\simeq 4\times 10^{-9}$. The corresponding saturation scale squared, $Q_s^2(x_2)$, will be $\sim  0.9$ GeV$^2$ (at $y=0$) and $\sim 10$ GeV$^2$ (at $y=\tilde{y}$). As $\mu^2\lesssim Q_s^2$, then $xG(y=0)\sim 3\sigma_0 Q_s^2/4\pi^2 \alpha_s \sim 5.4/\alpha_s$, $\sigma_{GGG}\sim 3\sigma_0/2$ (limit value deep in soft region) and $|\Psi_{GG}|^2\propto 2\alpha_s (N_c-1)/(8\pi)\delta (z-1/2)\delta(r^2-1/2m_G^2)$. This will give roughly,
\begin{eqnarray}
\frac{d\sigma_{tot}}{dy} (y=0) \sim \frac{3\sigma_0 Q_s^2(y=0)}{16\pi^3}\sigma_0\approx 12\,\mathrm{mb},
\end{eqnarray}
where in the simplified expression above the integration on rapidity is of order $2\times 1/\lambda\simeq 8.1$ (the rapidity dependence comes mostly from $Q_s^2(x_1=2m_Ge^y/\sqrt{s})\propto e^{-\lambda y}$). This would give quantitatively $\sigma_{tot}\sim 95$ mb, which is order of magnitude similar to measured cross section.

Finally, we have also considered another color dipole approach addressing the soft scattering processes. In such a case, other observables can be described as the elastic cross section and the elastic slope parameter.

\subsection{b-CGC and Eikonal models}
We follow Ref. \cite{Bartels:2002uf} and compute the total cross section in the following way, 
\begin{eqnarray}
\label{eq:N}
\sigma_{tot}^ {hp}(s) =  2 \int d^2b d^{2}rdz \left|\psi_h (r,z)  \right|^2\,N(s,r,b),
\end{eqnarray}
which depends on the color dipole amplitude, $N(s,r,b)$, and on the hadron wavefunction, $\Psi_h (r,z)$. The expression resembles the same equation for the DIS description within the color dipole approach. That is, the photon wavefunction is replaced by the hadron one. 
Furthermore, we consider the exponential approximation  of the elastic differential cross section at the diffraction peak,
\begin{equation}
\frac{d \sigma_{el}}{dt}\simeq\frac{d \sigma_{el}}{dt}\bigg |_{t=0} e^{B_{el}t}= \frac{\sigma_{tot}^{2}(1+\rho^{2})}{16\pi}e^{B_{el}t},
\label{eq:eldiff}
\end{equation}
where $t=-q_{t}^{2}$ is the momentum transfer in a $hp$ collision, $\rho$ is the real-to-imaginary ratio of the forward elastic amplitude
\begin{equation}
\rho^ {hp}(s) = \frac{\text{Re} A_{el}(s,t=0)}{\text{Im} A_{el}(s,t=0)}\simeq \frac{\pi}{2\sigma^ {hp}_{tot}}\frac{d\sigma^ {hp}_{tot}}{d\ln(s/s_{0})},
\label{eq:rho}
\end{equation}
and $B_{el}$ is the slope, which is given by $B_{el}(s)=B_{0}+B'(s)$ with $B_0=7.8 $ GeV$^{-2}$ and 

\begin{equation}
B'(s)=\frac{\int b^{2}d^{2}bd^{2}r  \ |\psi_{h}(r)|^2 N(r,b,x)}{\sigma_{tot}}=\frac{1}{2}\left\langle b^{2}\right\rangle \ .
\label{eq:Bel}
\end{equation}

In Eq.(\ref{eq:rho}) we invoke a first order Derivative Dispersion Relation (DDR) to provide an estimate of the parameter $\rho$ at LHC energies, especially at 13 TeV. Once the leading terms in the amplitude of dipole models are interpreted here in the Regge language as Pomeron terms (soft + hard), we have not accounted for Odderon signatures. Therefore, our predictions for $pp$ and $\bar{p}p$ observables are degenerate (the same being true for $\pi^{\pm}p$).

Finally, the elastic cross section can be computed by integrating eq. (\ref{eq:eldiff}) to give (as $\rho^{2}\ll 1$):
\begin{equation}
\sigma_{el}^{hp}(s)\simeq\frac{[\sigma_{tot}^{hp}(s)]^2}{16 \pi B_{el}^{hp}(s)}.
\end{equation}

Here, in the meson-proton scattering the meson is treated as a $q\bar{q}$ pair and calculations follows that of DIS, i.e., the interaction of a color dipole with a proton target and saturation physics can be embedded in the dipole amplitude. Similar approach has been considered also in Refs.\cite{Shoshi:2002in,Flensburg:2008ag}, where the Pomeron dynamics is written in terms of the dipole-dipole cross section. For instance, in Ref. \cite{Shoshi:2002in} the large dipoles are dominated by a soft Pomeron contribution whereas small dipoles are driven by a hard Pomeron. On the other hand, in Ref. \cite{Flensburg:2008ag}, based on the Mueller's cascade model the authors discuss several contributions including the effect of Pomeron loops.  

For the wave functions of mesons and baryons, we use the phenomenological ansatz by  Wirbel-Stech-Bauer (WSB) \cite{Shoshi:2002in}, which gives:
\begin{eqnarray}
\psi_h(z,r) = \sqrt{\frac{z(1-z)}{2 \pi S_h^{2} N_{h}}} \text{exp}\left(-\frac{(z-\frac{1}{2})^2}{4 \Delta z_{h}^{2}}-\frac{r^{2}}{4S_{h}^{2}}\right),
\label{eq:sh}
\end{eqnarray}
where the hadron wave function is normalized to unity
\begin{equation*}
\int \!\!dz d^2r \, |\psi_h(z,r)|^2 = 1.
\end{equation*}
This condition yields the following normalization constant, $N_{h}$:
\begin{eqnarray}
N_h = \int_{0}^{1} dz \ z (1-z) \ e^{-(z-\frac{1}{2})^2 / 2\Delta z_h^2}. 
\end{eqnarray}

Therefore, mesons and baryons are assumed to have a $q\bar{q}$ and
quark-diquark valence structure. As quark-diquark systems are equivalent to $q\bar{q}$ systems, this allows us to model not only mesons but
also baryons as color-dipoles. The values of parameters in our case are the following: $\Delta z_h =0.3\,(0.2)$ and $S_h = 0.86\,(0.607)$ fm, for $p/\bar{p}\,(\pi^ {\pm}$), respectively \cite{Shoshi:2002in}. $S_h$ is a fitted parameter which gives a measure of the transverse hadronic radius. Hence, as the hadron wave function has a Gaussian profile which is centered at $S_h$ (see Eq. (\ref{eq:sh})), it is expected that dipoles with approximately the hadron radius dominate the contribution to the cross sections.

At this point, some discussion is in order. In our calculations the proton is considered as a quark-diquark system having a meson-like structure. Specifically, the proton can be viewed as a bound state of an up-quark and an isospin zero, quark spin zero spatially extended 2-quark state, the diquark. In this case, quark-diquark systems are equivalent to quark-antiquark systems and we accordingly obtain $\langle z \rangle = 1/2$ from wavefunction for protons in Eq. (\ref{eq:sh}). The three-quark structure of a baryon makes the model to be notably complex  but produces similar phenomenological results as in the quark-diquark picture. A comparison between the three-body picture and the diquark one for protons concerning soft observables has been done in Ref. \cite{PhysRevD.50.1992} (see e.g. Fig. 1 and Table III in \cite{PhysRevD.50.1992}). On the other hand, in literature different treatments for baryon wavefunction are considered. For instance, in Ref. \cite{Nikolaev:1999qh} the proton wavefunction is obtained by the symmetric oscillator wavefunction of the valence three-quark proton. In this approximation, the proton is viewed as a 3/2 color dipoles spanned between quark pairs. The distribution of size of color dipoles with transverse size $r$ spanned between $q\bar{q}$ in the proton is considered gaussian, where $\langle r_p^2 \rangle=0.658$ fm$^2$. This value is not far from the value $S_h^2=0.740$ fm$^2$ appearing in Eq. (\ref{eq:sh}) for protons. Accordingly, the average $\langle z \rangle = 1/3$ is obtained. However, as we will see afterwards   the calculation of low mass single diffraction cross section becomes a hard task as 3 color centers from constituent quarks should be used and the average amplitude squared, $\langle N^2 \rangle$, has to be taken in each of these configurations.


Before discussing an impact-parameter dipole amplitude extracted from DIS data, we would  need to rewrite the energy dependence from photon-hadron scattering in terms of the appropriate Bjorken scaling variable-$x$. In this work, the following ansatz has been considered:
\begin{equation}
\frac{1}{x} = \frac{sr^2}{(s_{0}R_c^2)},
\label{eq:x_scm}
\end{equation}
which has been successfully considered in Ref. \cite{Donnachie:2001wt} and where $s_0^ 2 \sim m_h^2$ and $R_c = 0.2\,\mathrm{fm}$. Such an ansatz is numerically equivalent to the proposal $\frac{1}{x}=\frac{s}{Q_0^2}$, with $Q_0^2\sim (2m_q)^2\simeq m_h^2$, made in Ref. \cite{Bartels:2002uf}. For simplicity and faster numerical calculation we consider the last relation, where $Q_0^2$ is a free parameter to be fitted to the total cross section data, above cm energies $\sqrt{s}\gtrsim $ 100 GeV.

For the impact-parameter amplitude we first consider the parametrization based on the Color Glass Condensate ideas (called from now on b-CGC model). In the b-CGC, the color dipole-proton amplitude is given by, 
\begin{equation}
\label{eq:bcgc}
N (x, r, b)=\left\{\begin{array}{ll}
N_0\,\left( \frac{r Q_s}{2}\right)^{2\gamma_{eff}},& r Q_s\leq\,2 \\ 
1-\exp\left[ -\mathcal{A} \ln^2\left( \mathcal{B} r Q_s\right) \right], & rQ_s\,>\,2
\end{array}
\right.,
\end{equation} 
where the effective anomalous dimension and the saturation scale, $Q_s$, are defined as:  
\begin{eqnarray} 
\gamma_{eff} & = & \gamma_s\,\,+\,\,\frac{1}{\kappa \lambda Y}\ln\left(\frac{2}{r Q_s}\right), \\
 Q_s & = & \,\,\left( \frac{x_0}{x}\right)^{\frac{\lambda}{2}}\,\exp\left\{- \frac{b^2}{4\gamma_s B_{CGC}}\right\}\, ,
\end{eqnarray}
where, accordingly, $Y=\ln(1/x)$ and $\kappa=\chi''(\gamma_s)/\chi'(\gamma_s)=9.9$, with $\chi$ being the LO BFKL characteristic function. The updated values for the model's parameters are the following:  $B_{CGC}= 5.5\,\mathrm{GeV}^{-2}$, $\gamma_s =0.6492$, $ N_0=0.3658$, $x_0=6.9 \times 10^{-4}$ and $\lambda = 0.2023$, which have been published in Ref. \cite{Rezaeian:2013tka}. That fit was performed in the range $x\le 0.01$ and $Q^2\in [0.75, 650]\,\text{GeV}^2$, with $m_c=1.4$ GeV, using high precision combined HERA data. 


We have also tried an eikonal-like expression for the dipole amplitude, which has a different impact parameter dependence. The function $S(b)$ is now described by the dipole profile function. Namely, the amplitude has the following form:
\begin{eqnarray}
N(x,r,b) & = & 1-\exp\left(-\frac{1}{2}\hat{\sigma}(x,r)S(b) \right),
\label{eq:N_eik}
\end{eqnarray}
with
\begin{eqnarray}
\hat{\sigma}(x,r) &=& \sigma_0 \frac{(rQ_s(x))^2}{4},\\
S(b) &=& \frac{2\beta b}{\pi R^2}K_1(\beta b).
\label{eq:Sb_eik}
\end{eqnarray}

Moreover, we have considered the parameters for  $\hat{\sigma}$ from the GBW saturation model \cite{Golec-Biernat:2017lfv}, taking $R^2 = 4.5$ GeV$^{-2}$  and $\beta = \frac{\sqrt{8}}{R}$.

The eikonal-like model above is strongly inspired in the success obtained in Ref. \cite{Ben:2017xny}, where an universal expression of cross sections for the exclusive vector meson production and Deeply Virtual Compton Scattering (DVCS) in photon-proton and photon-nucleus interactions
based on the geometric scaling phenomenon has been obtained. Using the same form, Eq. (\ref{eq:N_eik}), it was found a theoretical parameterization based on the scaling property where cross sections depend only on the single variable $\tau = (\mu^2/Q_{s}^2)$ ($\mu^2=Q^2+m_V^2$ for vector mesons and $\mu^2=Q^2$ for DVCS, respectively). In that work, the saturation scale controls the energy dependence and nuclear effects, as well. The eikonal-like model then describes all available data from DESY-HERA for $\rho, \,\phi,\, J/\psi$ production and DVCS measurements. Furthermore, the photonuclear cross sections for $\rho$ and $J/\psi$ production extracted from the ultraperipheral heavy ion collisions at the LHC, i.e., $\sigma(\gamma Pb\rightarrow \mathrm{VM}+Pb$, are also quite well described.

\section{Fit results and Discussion}
\subsection{Total and Elastic Cross Sections}
\begin{figure}[h]
\hspace{-0.2cm} \includegraphics[scale=.48]{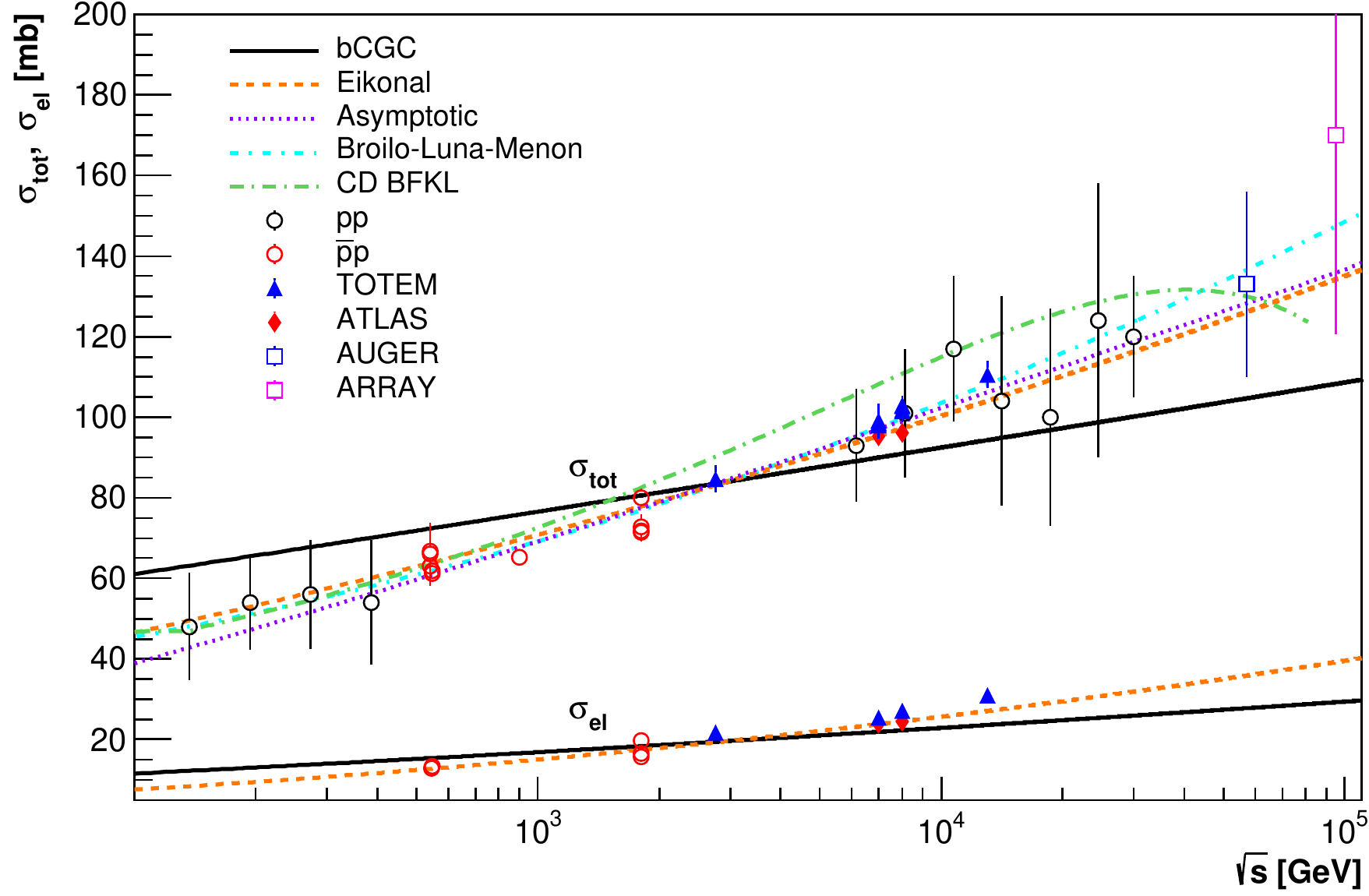}
\caption{The total and elastic cross sections for $pp$ collisions. The upper cross sections are total cross sections, while the lower ones are the elastic cross sections. Tevatron, SPS, LHC and cosmic rays data are presented \cite{Nemes:2018tk,Antchev:2017dia,Antchev:2017yns,Antchev:2016vpy,Aaboud:2016ijx,Aad:2014dca,Antchev:2015zza,Antchev:2013iaa,Collaboration:2012wt,Abbasi:2015fdr} The lines are results from models considered with fitted parameter $Q_0^2$.}
\label{fig:1}
\end{figure}

Fits to the $pp$ and $\bar{p}p$ total cross sections for the three models presented in last section are shown in Fig \ref{fig:1}. Both accelerator and cosmic rays data have been gathered from PDG2018 review \cite{Tanabashi:2018oca}, recent LHC measurements, mostly by  TOTEM and ATLAS Collaborations \cite{Nemes:2018tk,Antchev:2017dia,Antchev:2017yns,Antchev:2016vpy,Aaboud:2016ijx,Aad:2014dca,Antchev:2015zza,Antchev:2013iaa} as well as from Auger and Telescope Array Collaborations \cite{Collaboration:2012wt,Abbasi:2015fdr}. All fits have been performed using the TMINUIT class of the ROOT framework \cite{Brun:1997pa}, through the MIGRAD algorithm. Specifically, we minimize the total cross section data for $pp$, $\bar{p}p$ and $\pi^{+}p$ scatterings for $\sqrt{s}\geq 100$ GeV, using the chi-squared per degrees of freedom (d.o.f.), $\chi^{2}/d.o.f.$, criterium as a goodness of fit estimator. As previously mentioned, the Asymptotic model has only fixed parameters and for the b-CGC and the Eikonal models the only fit parameter to be tuned is $Q_0^2$. Best fit parameters of these models are thus given in Table \ref{tab:t1}.

\begin{table}[H]
 \centering
\caption{Best fit parameter $Q_0^2$ and chi-squared per degrees of freedom ($\chi^2 /$d.o.f.) of fits to $pp$ and $\pi^{+}p$ total cross section data for b-CGC and Eikonal models.}\label{tab:1}
\vspace*{.3cm}
\begin{tabular}{c|c|c}
\hline\hline
    Model & $Q_0^2$ \ [GeV$^2$] & $\chi^2 /$d.o.f. \\
\hline
 b-CGC ($pp$)   & $(9.44 \pm 0.57 )\times 10^{-5}$  & $518.40/22 = 23.6$  \\
\hline
 b-CGC ($\pi ^+ p$)   & $0.10  \pm 0.12 $  & $9.88/6 = 1.65$  \\
\hline
 Eikonal ($pp$) & $ 0.308\pm 0.019 $  & $70.25/22 = 3.19$   \\
\hline
 Eikonal ($\pi ^+ p$) & $13 \pm 14   $  & $9.25/6 = 1.54$   \\
\hline\hline
\end{tabular}
\label{tab:t1}
\end{table}

\begin{figure}[h]
\includegraphics[scale=.48]{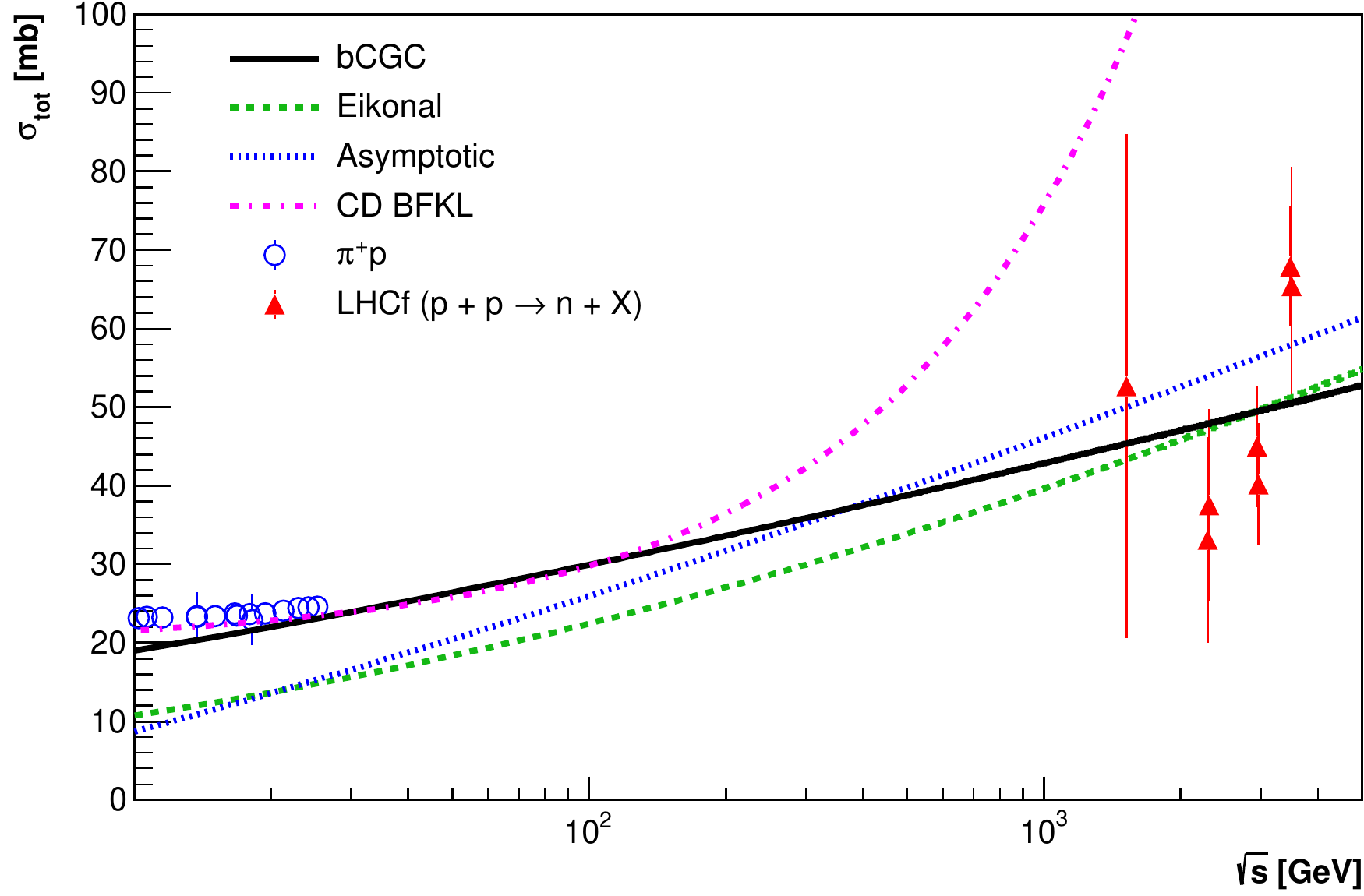}
\caption{ The total cross section for $\pi ^{+}p$ collisions. Data are obtained from inclusive leading neutrons spectra in the LHCf collaboration of the LHC \cite{Ryutin:2016hyi,Khoze:2017bgh}. Low energy data are also presented.}
\label{fig:2}
\end{figure}

As shown in Figs. \ref{fig:1} and \ref{fig:2}, the Asymptotic model provides a reasonable description of the data in the wide energy range, 100 GeV $<\sqrt{s}$ < 13 TeV. This  feature can be related to dominant role of gluon production at very low-x, as the model has only four fixed parameters, namely $m_{G},\ \sigma_{0}, x_{0}$ and $\lambda$.

On the other hand, the b-CGC model gives a slowly rising total cross section, with a pre-asymptotic form $\sigma_{tot}\sim a\ln s$. Such a behavior is related to the fact that dipoles with sizes nearly the proton radius dominate once the hadron wave function has a Gaussian profile centered at $R_p$. For $rQ_s \geq 2$, the dipole cross section in b-CGC is mainly driven by the Balitsky-Kovchegov (BK) equation assymptotic solution (see Eq. (\ref{eq:bcgc})). At high energies (small $x$), the saturation scale grows. Thereby, the quantity $rQ_s$ becomes larger and the dipole amplitude tends to the unity, which leads to the saturation regime. At LHC and cosmic rays energies, the system is saturated and the cross section has already reached its limit.

Some words of caution are in order at this point. Most of saturation models (b-CGC, Impact Parameter Saturation model IPSAT, GBW and so on) predict a  total cross section proportional to the quantity $\sigma_0\sim 27$ mb which is quite small compared to typical values of measured cross sections (even for the pion case).  This was discussed already in Ref.  \cite{Bartels:2002uf} for GBW model and the numerical solution of BK equation. The situation is similar here for the b-CGC model, where the smallness of overall normalization had to be compensated by an unrealistic value of the $Q_0^2$ parameter in $pp$ case (see Tab. \ref{tab:t1}).
The situation is different for the Eikonal model, where the overall normalization is given by the integration over impact parameter of the profile $S(b)\propto b K_1(\beta b)$, which corresponds to the proton dipole form factor in momentum transfer representation.

Regarding the Eikonal model, the profile function considered, $S(b) \sim \frac{b}{R} K_1 \big (\frac{b}{R} \big )$, results in a asymptotic total cross section, $\sigma_{tot}\sim \ln ^{2} s$, as long as $1/x\sim s$, as stated by the ansatz (\ref{eq:x_scm}) \cite{Bartels:2002uf}. This Froissart-bounded cross section can be naturally obtained in structure function models with leading asymptotic form $F_{2}(x)\sim \ln ^{2} (1/x)$, at low-$x$, such as in the model by Block-Durand-Ha \cite{Block:2013mia,Block:2013nia}, whose analytical form ultimately leads to the dipole cross section given in Ref. \cite{Arguelles:2015wba}. Thus, a deep link between dipole cross section (an its sizes) and the total cross section can not only be antecipated at very high-energies, but it is essential to understand low-$x$ parton dynamics. In addition, the same asymptotic energy behavior is seen in soft Pomeron models, such as BLM \cite{Broilo:2018qqs}, COMPETE \cite{Cudell:2002xe} and PDG \cite{Tanabashi:2018oca}. 

To stablish a direct comparison with other popular models of current Regge phenomenology we also plot in Fig. \ref{fig:1} the  prediction of the model by Broilo-Luna-Menon (BLM) \cite{Broilo:2018qqs}, in which the energy dependence of the soft Pomeron is parametrized as follows  (Modell II):
\begin{equation}
\sigma_{tot}^{\mathbb{P}}(s)=A+D \ln^2 (s/s_0) , 
\end{equation}
where $A=29.6 \pm 1.2 \ mb$, $D=0.251 \pm 0.010 \ mb$ and $s_{0}=4m_{p}^{2}\simeq 3.521$ GeV$^{2}$. As this model is inspired in the COMPETE analysis (pre-LHC) \cite{Cudell:2001pn,Cudell:2002sy,Cudell:2002xe} we shall refer to it as ``Broilo-Luna-Menon (BLM)'' model.

We have also estimated the pion-proton total cross section. Our predictions are shown in Fig. \ref{fig:2} compared to recently extracted data from leading neutron production in the TeV region \cite{Ryutin:2016hyi,Khoze:2017bgh} by using recent data from the LHCf Collaboration \cite{Adriani:2015nwa} The magnitude and energy evolution predicted by the models tested is in quite good agreement with the data, despite their large error bars.

For the Asymptotic model we use the additive quark model, where $\sigma_{tot}^{\pi N}/\sigma_{tot}^{NN}=2/3$. Concerning the b-CGC and Eikonal models, we explicitly take into account $|\psi_{\pi}(z,r)|^2$ from Eq. (\ref{eq:sh}). The low energy data are also presented. In the models discussed so far only the Pomeron contribution is being computed. For low energy a nonperturbative contribution as well as the Reggeon piece have to be added.

Predictions of models for the total cross section at LHC energies of $7 \ TeV$, $8 \ TeV$, $13 \ TeV$ and $14 \ TeV$ and at the cosmic ray energies, $57 \ TeV$ (Pierre Auger Observatory) and $95 \ TeV$ (Telescope Array), are shown in Table \ref{tab:2}. It is important to mention that we have not presented the b-CGC predictions because it did not have a good agreement with data, as it can be clearly seen in Fig \ref{fig:1}. Thereby, for the observables calculated in the next sections, we will not take into account the results presented by this model.

\begin{table}[h]
\caption{Predictions of $\sigma_{tot}^{pp/\bar{p}p}$ for the Asymptotic and Eikonal models.}\label{tab:2}
\begin{tabular}{c|c|c}
 \hline\hline 
 $\sqrt{s}$ (TeV)   & Asymptotic - $\sigma_{tot}^{pp/\bar{p}p}$ (mb) & Eikonal - $\sigma_{tot}^{pp/\bar{p}p}$ (mb)  \\
 \hline
  $7.0 $   & $96.9$      & $95.4$  \\
  \hline
  $8.0 $   &$99.0$      & $97.2$   \\
  \hline
  $13 $  & $106$     & $104$ \\
  \hline
  $14 $  & $107$     & $105$  \\
  \hline
  $57 $  & $128$    & $126$  \\
  \hline
  $95 $  & $136$   & $134$  \\
 
 \hline\hline
\end{tabular}
\end{table}

Before analysing the $\rho$-parameter and hadronic forward slope in next subsection,  we explicitly compare our predictions to the color dipole BFKL-Regge expansion (CD-BFKL) approach by Fiore et al. (Ref. \cite{{Fiore:2012yi}} and references therein). The main ingredient in this formalism is the BFKL dipole cross section, $\sigma (Y=\ln(x_0/x),r)$, which sums the $\sim \alpha_s\ln (1/x)$ multi-gluon production cross sections in perturbative-QCD. The initial condition for the evolution at $x = x_0$ and dipoles having transverse size $r$ is the Yukawa screened two-gluon exchange. The evolution equation for the dipole cross section concerning the non-unitarized running CD BFKL amplitudes is given by,
\begin{eqnarray}
\frac{\partial \,\sigma (Y,r)}{\partial Y} & = & \int d^2\vec{r}_1 \left| \psi(\vec{r}_1)- \psi(\vec{r}_2)\right|^2 \nonumber \\
&\times & \left[\sigma_3(Y,\vec{r},\vec{r}_1, \vec{r}_2)-\sigma (Y,r)\right]
\end{eqnarray}
where $\psi (\vec{r})\propto \frac{\hat{r}}{R_c}K_1(r/R_c) $ is the radial light-cone wavefunction of the dipole with the vacuum screening of infrared gluons (infrared cutoff regulator is $R_c=0.26$ fm). The $q\bar{q}g-$nucleon three-parton cross section, Eq. (\ref{eq:qqG}), is a function of $\vec{r}_{1,2}$, which are respectively the quark-gluon and antiquark-gluon transverse separations in the two-dimensional impact parameter plane for dipoles generated by the quark-antiquark color dipole source. 

In \cite{{Fiore:2012yi}}, the unitarity absorption corrections are computed using the BK non-linear BFKL equation in the impact parameter representation. The evolution equation in this case reads as,
\begin{eqnarray}
\frac{\partial \,\sigma (Y,r)}{\partial Y} & = & \int d^2\vec{r}_1 \left| \psi(\vec{r}_2)- \psi(\vec{r}_1)\right|^2 \nonumber \\
&\times & \left\{\sigma(Y,\vec{r}_1)+\sigma (Y,\vec{r}_2)-\sigma (Y,r)\right.\nonumber \\
&-& \left. \frac{\sigma(Y,\vec{r}_1)\sigma(Y,\vec{r}_2)}{4\pi\,B_{12}}\exp\left[ -\frac{r^2}{8B_{12}} \right] \right\},
\end{eqnarray}
where $B_{12}=B_1(Y,r_1)+B_2(Y,r_2)$ and $B_i=B(Y,r_i)$. The authors consider that the  diffraction cone slope, $B$, drives the area populated with interacting gluons. Specifically, the diffraction slope for the forward cone in the dipole–nucleon scattering is given by the expression  $B(Y,r)=(r^2/8)+(R_N^2/3)+2\alpha_{\pom}^{\prime}Y$, with $R_N^2\simeq 12$ GeV$^{-2}$ and $\alpha_{\pom}^{\prime}\approx 0.1$ GeV$^{-2}$.

Accordingly, in Fig. \ref{fig:1} it is shown the predictions of Fiore at al. \cite{Fiore:2014hga} for the total cross section (green dot-dashed curve). We see that absorptive corrections are strong at cosmic rays interaction  and at the highest collision of the LHC. Up to 2 TeV their results are very similar to ours. For completeness, we also add the predictions from the CD BFKL approach (without absorption corrections) for the pion-proton cross section in Fig. \ref{fig:2}. The dot-dashed curve gives the hard contribution to $\pi N$ cross section taken from Ref. \cite{Nikolaev:1999qh} (Eq. (28) and parameters in Table 1 from that reference). The low energy data is nicely described and the high energy LHC values could be reproduced in case absorption is included. The absorption effect should be similar to the proton case in the same energy range.

\subsection{Real-to-imaginary ratio and the forward slope}

\begin{figure}[h]
\includegraphics[scale=.48]{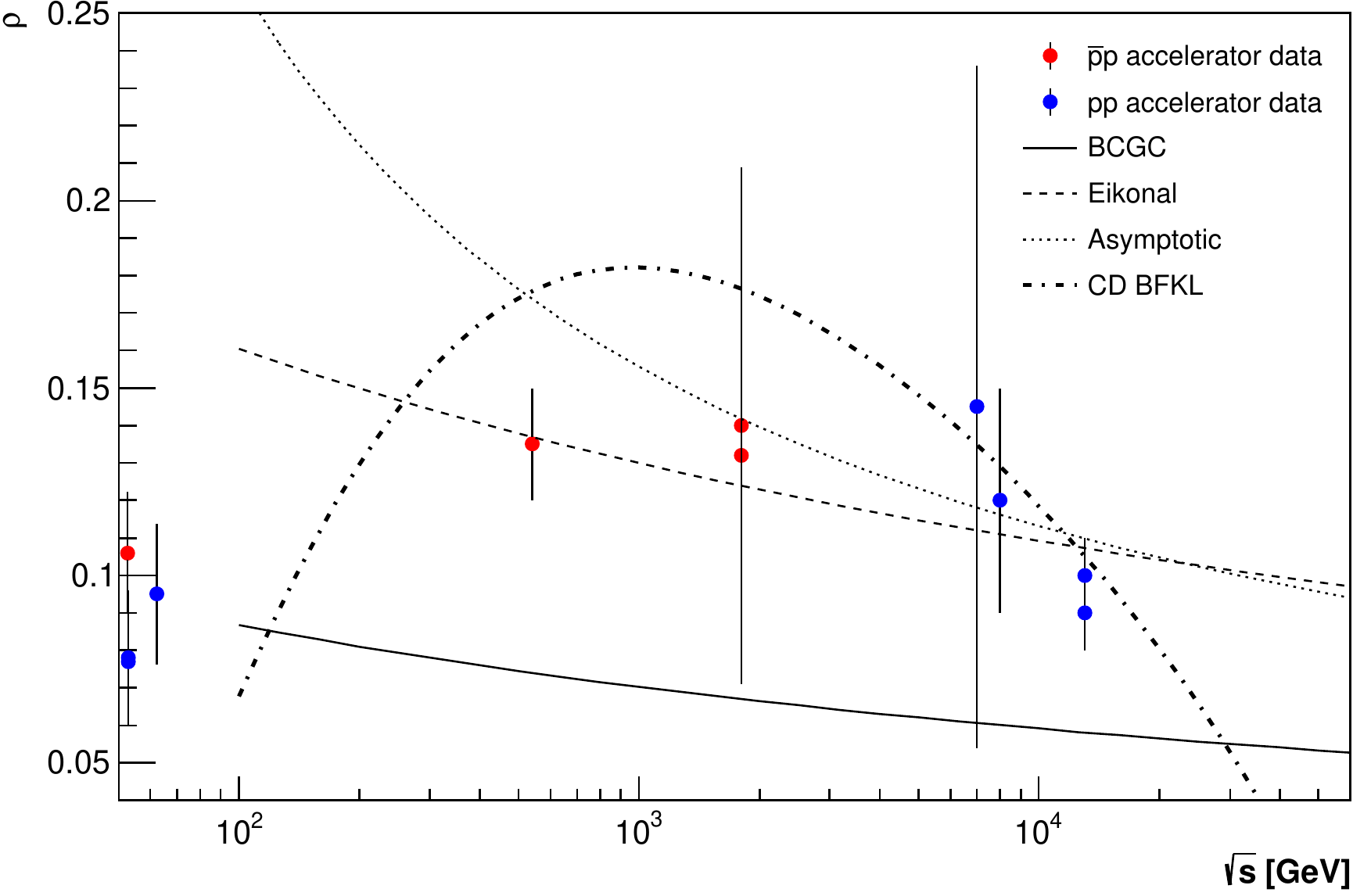}
\caption{Real-to-imaginary ratio predictions of models b-CGC, Eikonal, Asymptotic and prediction from CD BFKL approach together with recent LHC data \cite{Antchev:2017yns}.}
\label{fig:3}
\end{figure}

\begin{figure}[h]
\includegraphics[scale=.48]{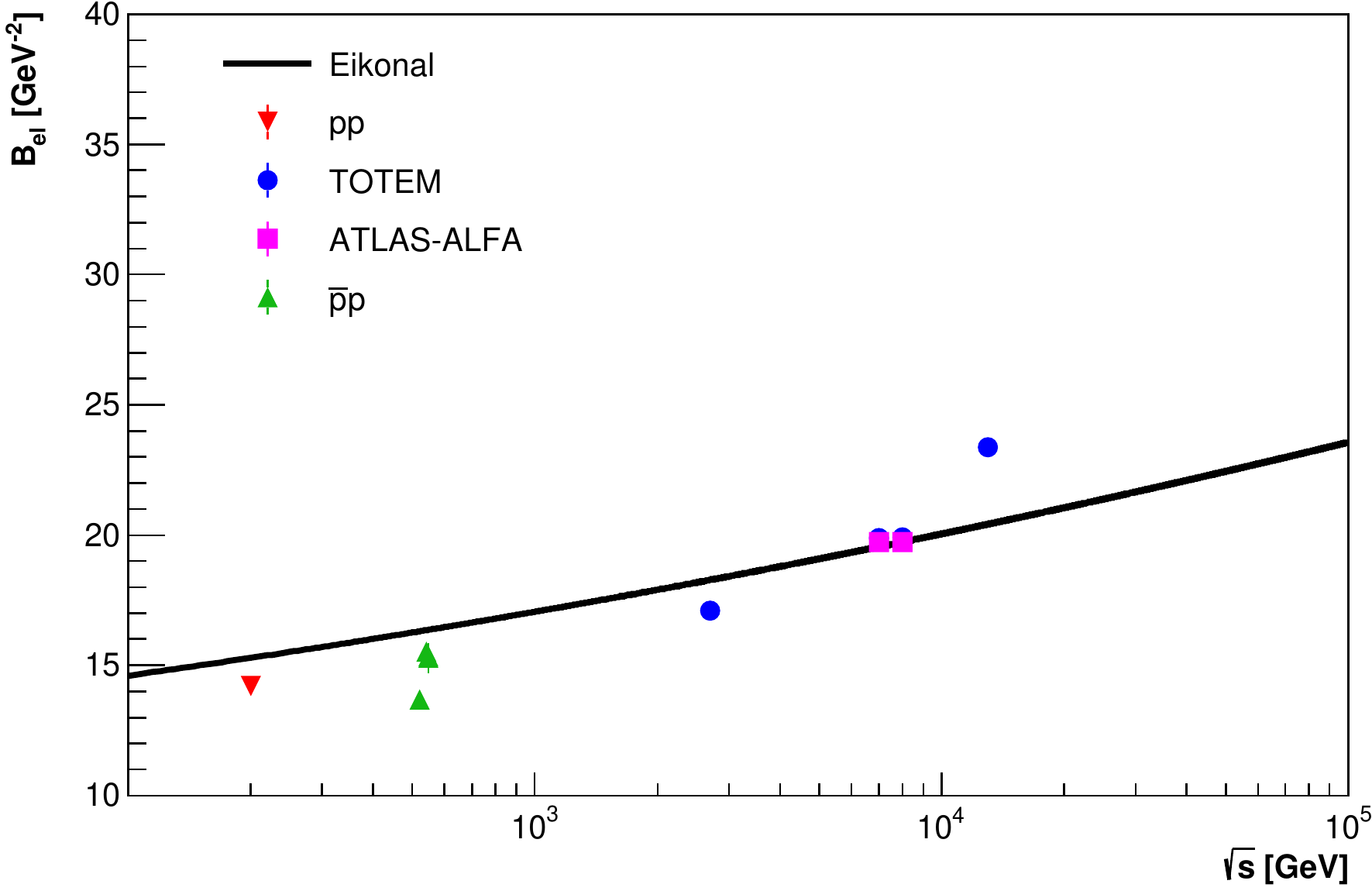}
\caption{Energy dependence of the slope, $B_{el}^{pp}$, predicted by the Eikonal model.}
\label{fig:4}
\end{figure}

 Furthermore, we also give predictions for two forward energy-dependent observables: (i) $\rho^{pp, \bar{pp}}(s)$, the real-to-imaginary ratio of the elastic amplitude, which follows in Fig. \ref{fig:3} and (ii) $B^{pp, \bar{pp}}_{el}(s,t=0)$  the forward slope, which is shown in Fig. \ref{fig:4}. Both plots comprise very recent LHC data and especially for $\rho$, an adequate description of the LHC13 datum (within error bars) is achieved. On the one hand, predictions from dipole models deviate significantly from the data, especially the b-CGC model. Such behavior is related to the very rapid decrease of the $b$ distribution at large impact parameters, $N(r,x,b)\sim 1-\exp^{-\alpha(r,x) b^{4}}$, which approximately follows a black-disk shape, $N(r,x,b)\sim \Theta(b-R)$, and leads to an almost flat energy dependence of $B_{el}$. We recall that such behavior is very similar to those presented in Ref. \cite{Bartels:2002uf}, where the GBW and Glauber-Mueller models for dipole cross section were considered. In Ref. \cite{Fiore:2014hga} no prediction is done for the forward slope, however an estimation can be done using the forward cone in the dipole–nucleon scattering. Taking for simplicity that the average dipole size is $\langle r \rangle \approx \sqrt{\langle r^2_p\rangle }$ the corresponding slope is $B_{dip}(s,\langle r \rangle )\approx (\langle r_p^2\rangle/8)+(R_N^2/3)+2\alpha_{\pom}^{\prime}\ln(sx_0/m_{\rho}^2)$.  At 7 TeV a rough estimation is $B_{el}=B_0+B_{dip}\approx (7.8  + 8.87)\,\mathrm{GeV}^{-2}\simeq 17\, \mathrm{GeV}^{-2}$ which is close to TOTEM measurement ($B_{el}^{exp}(7\,\mathrm{TeV})=19.9\pm 0.3\,\mathrm{GeV}^{-2}$).

\begin{figure}[h]
\includegraphics[scale=0.48]{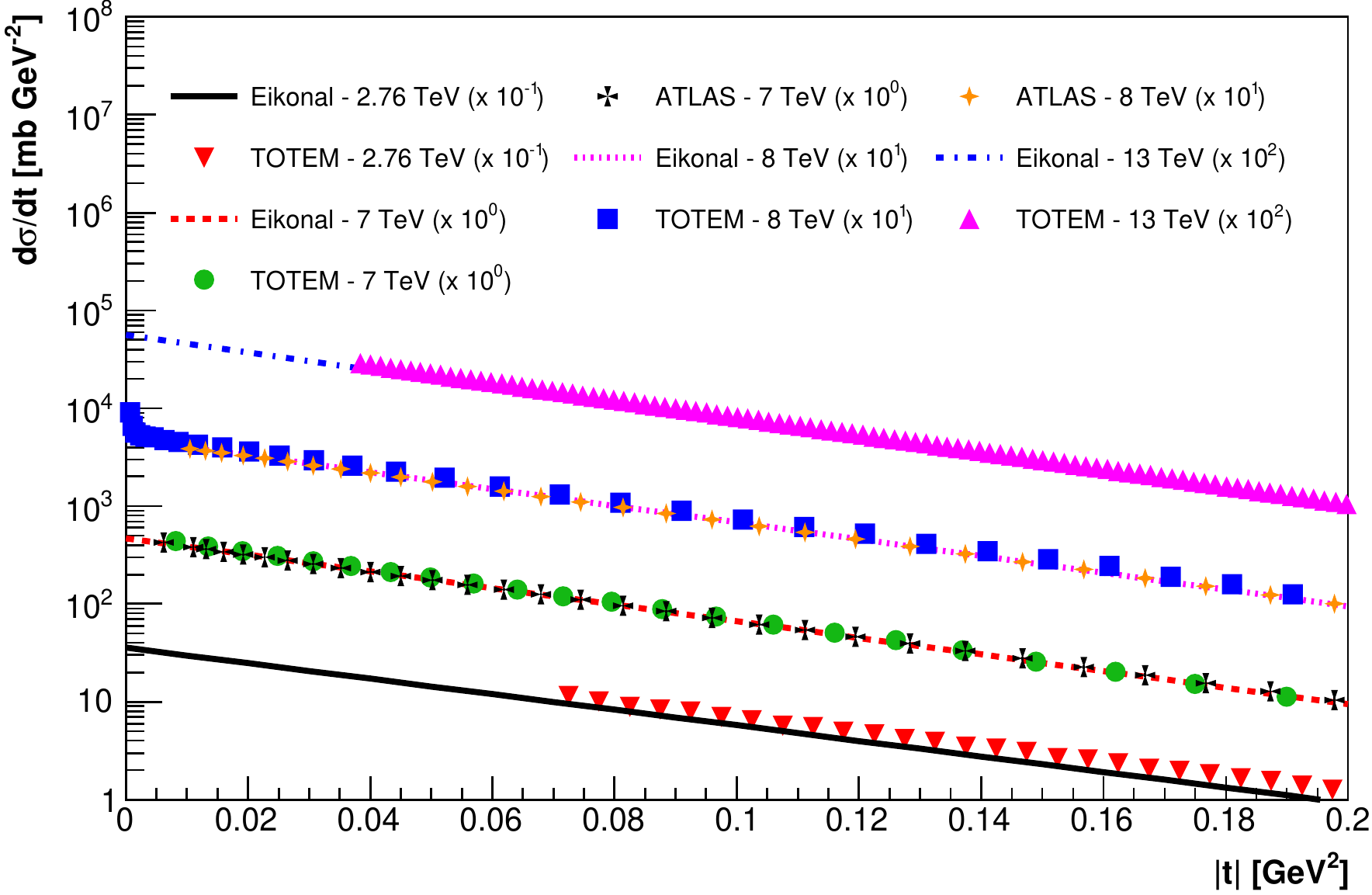}
\caption{Differential elastic cross section data measured by the ATLAS and TOTEM Collaborations at $\sqrt{s}=2.76 \ TeV$ \cite{Antchev:2018rec}, $7 \ TeV$ \cite{Aad:2014dca,Antchev:2013gaa}, $8 \ TeV$ \cite{Aaboud:2016ijx,Antchev:2016vpy} and $13 \ TeV$ \cite{Antchev:2018edk} and predictions of the Eikonal model in the region $0<-t\lesssim 0.2$ GeV$^{2}$.}
\label{fig:5}
\end{figure}

Conversely, due to smoother $b-$distribution given in Eqs. (\ref{eq:N_eik}) and (\ref{eq:Sb_eik}), the Eikonal model shows a better agreement with data, yielding a more acceptable trend of rising for $B_{el}(s)$. Indeed, as we show in Fig. \ref{fig:5}, predictions of this model for the elastic differential cross section reproduce the general structure of the diffraction cone ($0<-t\lesssim 0.2$ GeV$^{2}$) at LHC energies, especially at 7.0 and 8.0 TeV.

Finally, we present the corresponding $\rho$-parameter for CD BFKL model  based on the results for the total $pp$ cross section and making use of dispersion relations (DDR). It is shown in Fig. \ref{fig:3} (black dot-dashed curve), which is driven by the change of inflexion in the total cross section in high energy collider region. The normalization is still in agreement with LHC data, whereas the shape shows somewhat a disagreement.

\subsection{Low mass diffraction}

For incorporating color transparency in a natural way, color dipole models are a perfect framework to study inelastic diffraction. Indeed, color dipoles can be regarded as eigenstates of diffraction \cite{Kopeliovich:2006fp}. 

\begin{figure}[t]
\includegraphics[scale=.48]{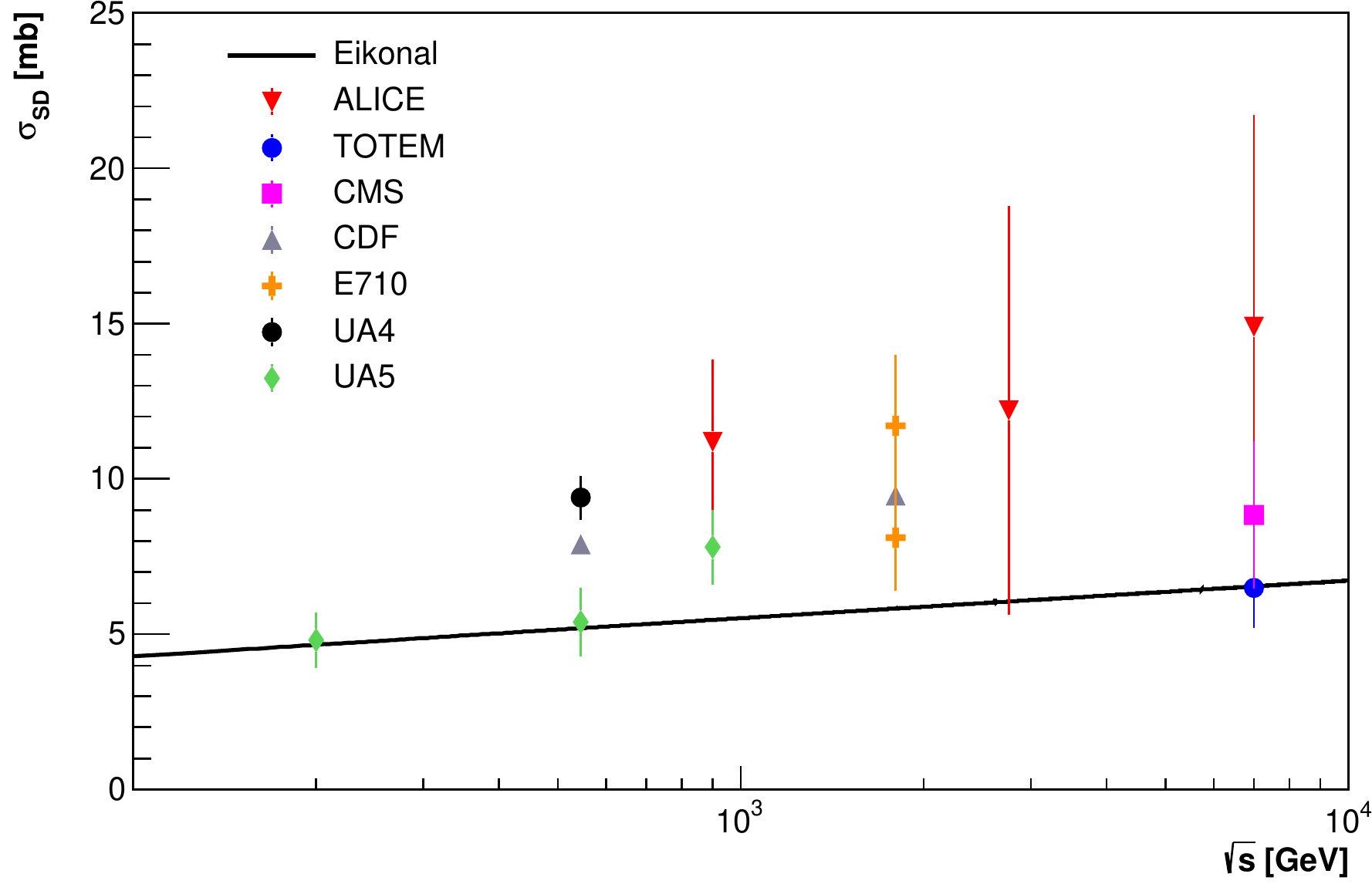}
\caption{Single diffractive (SD) cross section for the reaction $pp/\bar{p}\rightarrow pX$ as a function of centre-of-mass energy, $\sqrt{s}$. The curve is the result for the color dipole amplitude from Eikonal model and consider only low mass diffraction contribution.}
\label{fig:6}
\end{figure}

In the one-channel models we have developed so far, low mass inelastic diffractive eigenstates can be treated using the Good-Walker (GW) mechanism \cite{Good:1960ba}. Since diffraction arises from fluctuations in high-energy scattering amplitude, we calculate the contribution of color dipoles to the single diffractive cross section in the low  mass (LM) region through the following relation:
\begin{eqnarray}
\sigma_{\mathrm{SD}}^{\mathrm{LM}}(s) =  \left\langle N^{2}\right\rangle -   \left\langle N\right\rangle^{2},
\label{eq:sdxsect}
\end{eqnarray}
where
\begin{align*}
\left\langle N^{2}\right\rangle &=\left(\int d^2b \int dz d^2r \left|\Psi_h (r,z)  \right|^2\,N^2\right), \\
\left\langle N\right\rangle &=  \int d^2b \left( \int dz d^2r \left|\Psi_h (r,z)  \right|^2\,N \right).
\end{align*}
The first term in Eq. (\ref{eq:sdxsect}) encompasses the quasi-elastic cross section term, where excitations of the target (beam) particle can occur in the interaction with dipoles within the proton. The second term corresponds to the pure elastic scattering term. The predictions for the dipole model are presented in Fig. \ref{fig:6}, as a function of the center-of-mass energy.  The theoretical curve (we choose the Eikonal model as reference) is compared to non-LHC collider data (ISR \cite{Armitage:1981zp}, UA4 \cite{Bernard:1986yh}, UA5 \cite{Alner:1987wb}, E710 \cite{Amos:1992jw} and CDF \cite{Abe:1993wu}) and the recent LHC measurements. In particular, we consider the  ALICE data \cite{Abelev:2012sea} at $\sqrt{s}=0.9, \,2.76$ and 7 TeV ($M_X<200$ GeV/c$^2$), the measurements of TOTEM \cite{BERRETTI:2013vra} ($3.4<M_X<1100$ GeV/c$^2$) and CMS \cite{CIESIELSKI:2013ksa} ($12<M_X<394$ GeV/c$^2$), as well. Similar approach as ours is presented in Ref. \cite{Flensburg:2010kq}, where fluctuations in the BFKL ladder are taken into account. It was demonstrated that in high energy proton-proton collisions these fluctuations are strongly suppressed by parton saturation.


The Good-Walker formalism was originally conceived so as to describe a system of a nucleon plus its diffractive $N^*$ isobars. Clearly, this simplistic approach is not suitable for high energy diffraction where $M_{diff}^2$ is bounded by $0.05s$, leading to a continua of diffractive Fock states \cite{Maor:2009tc}. 

GW models shortcomings are amended once multi Pomeron interations are included, leading to a high mass diffraction \cite{Maor:2008tf}. If we consider a single diffractive channel $p+p \to p+ M_{SD}$, Mueller's triple Pomeron mechanism yields high SD mass which in non GW \cite{Maor:2009tc}. 


CDF analysis suggests a relativily large value for $G_{3 I\!P}$ (see Ref. \cite{Maor:2009tc}). Therefore, it is necessary to consider a very large family of multi Pomeron interactions (enhanced $I\!P$) which are not included in the GW formalism. This dynamical feature becomes significant above Tevatron energy and leads to profound differences in the calculated values of soft cross sections. It can be seen in Fig. \ref{fig:6} that the results from Eq. (\ref{eq:sdxsect}) do not show a good agreement with data at high energies due to the fact that high mass diffraction is not taken into account. 

Despite high mass diffractive dissociation is out of scope for the present study it can be properly addressed in a color dipole approach. For example, as referred before both low and high mass excitation was described by the Good–Walker mechanism in Ref. \cite{Flensburg:2010kq}. In that work, the high mass diffraction is connected to fluctuations in the BFKL evolution and it is shown that in $pp$ collisions unitarity constraints and saturation decrease those fluctuations towards the black disc limit of scattering process. Moreover, the Dipole Cascade Model can reproduce the expected triple-Regge form for the bare pomeron with $\alpha_{\pom}(0)=1.21$ and $\alpha_{\pom}^{\prime}=0.2$ GeV$^{-2}$, and the triple-pomeron coupling is shown to be almost constant, $g_{3\pom}\approx 0.3$  GeV$^{-1}$.  It is argued that GW and triple-pomeron formalisms for high mass dissociative diffraction are just different  aspects of the same phenomenon. Specifically, in both approaches diffractive excitation is the shadow of absorption into inelastic channels. This conclusion is not completely new as in the seminal work in Ref. \cite{Genovese:1994wy} where a direct computation of the triple-pomeron coupling for both  diffractive photoproduction and DIS at large-$Q^2$ has been done within the CD BFKL formalism already discussed. It was found a weak dependence on $Q^2$, producing $G_{3\pom}(Q^2)\approx 0.23$ GeV$^{-2}$ at $Q^2=0$ and $G_{3\pom}(Q^2)\approx 0.36$ GeV$^{-2}$ for $Q^2\geq 3$ GeV$^2$. In the context of the formulations presented here, within the color dipole picture the high mass dissociation can be understood as a three stage process. First, the penetration of the projectile dipole through the target without inelastic interaction then followed by the the emission of one extra gluon (considered a new dipole in large $N_c$ limit). Finally, one has the interaction of two produced dipoles with the target. The main ingredient in last stage is the amplitude of gluon-dipole scattering that has been investigated in Ref. \cite{PhysRevD.65.074026}. Starting from the dipole amplitude in Eq. (\ref{eq:N}) written in terms of opacity function $\Omega$, one has  $N(s,r,b)=1-\exp[-\frac{1}{2}\Omega(s,r,b)]$. For instance, in our eikonal-type model, Eq. (\ref{eq:N_eik}), $\Omega(s,r,b)= \hat{\sigma}(x,r)S(b)$. For the proton considered as an effective color-dipole (the quark-diquark picture) the high mass diffraction cross section reads as:
\begin{eqnarray}
M^2\frac{\sigma_{\mathrm{SD}}^{\mathrm{HM}}(s)}{dM^2} & =& \frac{\alpha_s N_c}{2\pi }\int d^2b \,dz\, d^2r |\psi_p(r,z)|^2\,e^{-\Omega(s,r,b)} \nonumber \\
&\times & \left[e^{-\Omega(s,r,b)}r^2I_1(\frac{s}{M^2},r,b)-r^2I_2(s,r,b) \right], \nonumber \\
\end{eqnarray}
with the following auxiliary integrals,
\begin{eqnarray}
I_1 &=&\int_{r^2}^{\infty} \frac{dr^{\prime 2}}{r^{\prime 4}}\left\{ 1-\exp\left[-\left(\Omega(\frac{s}{M^2},r^{\prime},b)-\frac{1}{2} \Omega(\frac{s}{M^2},r,b) \right)  \right] \right\}^2, \nonumber \\
I_2 &=& \int \frac{d^2r^{\prime}}{2\pi r^{\prime 2}(\vec{r}-\vec{r}^{\prime})^2}\,\left[1-\exp\left(-\Omega(s,r,b)  \right)  \right]^2,
\end{eqnarray}
where $I_2$ is related to a change for elastic scattering of the original dipole (with transverse size $r$) due to emission of an extra gluon. In $I_1$ the expression in curly bracket is the amplitude for gluon-dipole scattering \cite{PhysRevD.65.074026}. Applications of the above formalism to the $pp$ scattering  will be postpone for future studies.

As a final comment on the expression, Eq. (\ref{eq:sdxsect}), for the low-mass contribution to the SD cross section we see it is suitable for computing the corresponding proton-nucleus ($pA$) cross section. This can be performed by replacing the proton profile function $S(b)$ in our case by that one extracted from nuclear form-factors, $S_A(b)$ (Woods-Saxon or similar parametrizations). The investigation about the size of nuclear effects in single diffraction is an open question in literature. For instance, in Ref. \cite{Santos:2014tka} predictions for SD cross section in $pPb$ collisions at the LHC are obtained in the context of Glauber model for nuclear scatterings and taking into account Regge phenomenology (including an effective Pomeron flux, which describes the measured SD cross section in $pp$ collisions). Recently,  in Ref. \cite{Goncalves:2019agu} the authors investigate the diffractive excitation in $pA$ collisions based on the dynamics of relativistic nuclear collisions through the concept of hadronic cross-section fluctuations. These fluctuations are related to inelastic shadowing and diffractive dissociation and their effect decreases at larger energies and heavier nuclei.

\section{Summary}
\label{summary}

In summary, we have applied to soft hadron-hadron scattering the color dipole picture including the parton saturation phenomenon as the transition region between soft and hard domain. We have shown that the inclusive process is mainly driven for dipole sizes near the saturation radius in the high energy regime. The main advantage is that the corresponding phenomenology is almost free of parameter as they are completely constrained from DIS data in $ep$ interactions. The models rely on the dipole cross section or $b$-dependent dipole amplitude and indicate that the impact parameter profile is crucial for a good data description. In this context, our best results followed from the Eikonal model, for which a smoother impact parameter structure was built. In fact, the wealth of high energy elastic scattering data can be nicely described by this model, including  $\sigma_{tot}$, $\sigma_{el}$, $\rho$, $d\sigma_{el}/dt$ in the diffraction cone and $\sigma_{SD}$ in the low mass region, using an one-channel eikonal approach. These findings indicate a possible path of exploring even further the color dipole formalism as an alternative approach to the more tradicional Regge-Pomeron calculus to handle soft hadron-hadron and hadron-nucleus scattering processes, where, for instance, the role of multiple parton interactions can be properly addressed. We are currently investigating this possibility.

\section*{Acknowledgments}

This work was  partially financed by the Brazilian funding
agencies CNPq and  CAPES. DAF acknowledges the support by the project INCT-FNA (464898/2014-5).

\bibliographystyle{h-physrev}
\bibliography{Soft_dipolesV6}

\end{document}